\documentclass[12pt]{article}

\usepackage{slashbox}
\usepackage{epsfig}

\begin{document}

\begin{titlepage}

\hfill\parbox{4cm} { }

\vspace{25mm}

\begin{center}
{\Large \bf Domain wall cosmology and multiple accelerations}

\vspace{10mm}
$\textrm{Bum-Hoon Lee}^{\dag,\ddag}$\footnote{bhl@sogang.ac.kr},
$\textrm{Wonwoo Lee}^{\dag}$\footnote{warrior@sogang.ac.kr},
$\textrm{Siyoung Nam}^{\dag}$\footnote{stringphy@gmail.com}, and
$\textrm{Chanyong Park}^{\dag}$\footnote{cyong21@sogang.ac.kr}
\\[15mm]

$^\dag$ \textit{CQUeST, Sogang University, Seoul, Korea 121-742} \\[5mm]
$^\ddag$ \textit{Department of Physics, Sogang University, Seoul, Korea 121-742}\\[5mm]

\end{center}

\thispagestyle{empty}

\vskip1.5cm


We classify the cosmological behaviors of the domain wall
under junctions between two spacetimes in terms of
various parameters: cosmological constants of bulk
spacetime, a tension of a domain wall, and mass parameters
of the black hole-type metric. Especially, we consider the
false-true vacuum type junctions and the domain wall
connecting between an inner AdS space and an outer AdS
Reissner-Nordstr${\rm \ddot{o}}$m black hole. We find
that there exist a solution to the junction equations
with an inflation at earlier times and an accelerating
expansion at later times.

\vspace{1cm}


\end{titlepage}

\newpage
\setcounter{footnote}{0}
\section{Introduction}
\noindent The recent cosmological observations of high red
shift supernova\cite{redshift} and the measurements of
angular fluctuations of cosmic microwave background
fluctuations \cite{cmb} imply that our present universe
expands with {\it acceleration}, instead of deceleration.
Thus, a theory describing our universe must incorporate in
it the accelerating era at present as well as the early inflationary era. It is very interesting to construct a model
explaining these facts.

The setups of the so-called bubble dynamics have opened the
possibilities to search the cosmology on the bubble using
the Israel junction equation. However, the junction
equations only describe the evolution of the vacuum bubble
and can not tell us about how to generate bubble. The
possible mechanisms of bubble formations were studied ini the case of
the true vacuum bubble case\cite{coleman1,coleman2,parke}
and the false vacuum bubble\cite{lee2,kim1,hackworth,lee1}. In \cite{farhi1,
blau1}, the dS-S junction, the wall connecting the false
vacuum inside it and the true vacuum outside it, was
researched. Similarly, junctions among spacetimes with
different cosmological constant, or vacuum energy, were
also handled. See dS-SdS\cite{aguirre1},
dS-SAdS\cite{freivogel1}, dS-RN AdS\cite{alberghi1} and
references therein\footnote{The inner space is written  in
the left and the outer in the right like
dS(``inner'')-SdS(``outer''). S, dS, AdS, and RN stand for
Schwarzschild, de Sitter , Anti-de Sitter, and
Reissner-Nordstr${\rm \ddot{o}}$m, respectively. For
example, the notation dS-SdS means that the space inside
the wall is described by the de Sitter-type metric and the
space outside the wall by de Sitter Schwarzschild black
hole-type metric.}. However, the main interest of them was
the observations in the bulk sides.

Recently, Randall and Sundrum (RS)
\cite{Randall:1999ee,Randall:1999vf} made an interesting
proposal that we may live in a four dimensional domain wall
(or brane) of the five dimensional AdS spacetime. In this
model, the metric fluctuation around the domain wall
admits a bound state and the background cosmological
constant makes this bound state a zero-mode to be
identified as a graviton. The tension of the wall is fine
tuned so that the effective cosmological constant on the
domain wall is zero and becomes stationary.  If the tension
of wall is not fine tuned, it was shown that the domain
wall can move in the five dimensional background space,
which causes the inflationary cosmology on the domain wall
\cite{Park:2000ga}. Kraus pointed out that this inflation
can be interpreted as the motion of the domain wall in a
stationary background\cite{Kraus:1999it}. When two pieces
of AdS spaces are glued along the moving domain wall, the
junction equation determines the motion of the domain wall
in terms of the domain wall tension and the two bulk
cosmological constants.

When considering the false vacuum bubble solution, since an
observer living in the outside space feels the positive
energy lump at center, the outside space can be considered
as a black hole-type metric by Birkhoff's theorem\cite{Birkhoff}. From now
on, we set the outside metric as a black hole type which
gives rise to the maximally symmetric space turning off the
black quantities like mass, charge and angular momentum.

In section 2, we will shortly review the junction equation.
In section 3, when the outside space is a Schwarzschild
black hole type metric, we will fully classify the
cosmology of the domain wall according to the parameters
which gives the same the qualitative behavior of the domain
wall cosmology investigated by many authors. In section 4,
when considering the outside space as a RN black hole type
we will show that there can exist a multi-inflation
solution, that describe the early time inflation as well as
the late time acceleration. This solution can describe
inflation of the early universe and of the present universe
at the same time with the FRW expansion in the middle
period. Here, though we find only one special solution
showing the multi-accelerating behavior, it is also very
interesting to find a parameter region describing the real
world comparing with many cosmological data. In section 5,
we close this paper with some discussion.


\section{Junction Equations in Einstein Frame}
In this section we will briefly review the method of
deriving Israel's junction equations from the variational
principle. The reader who is familiar with the junction equation may skip this
section. Let $M$ be a (D+1) dimensional manifold which a
domain wall $\Sigma$ partitions into two parts $M_{\pm}$.
We shall denote the two sides of $\Sigma$ as
$\Sigma_{\pm}$, which are boundaries of both bulk
$M_{\pm}$. We assume that each metric in the bulk $M_\pm$
is well defined solution of Einstein's equation in the sense
that we seek the whole solution of Einstein equation in the
manifold $M$ as a distribute type. We also demand that the
first junction condition should be satisfied so that across $\Sigma$ the tangential components of the metric
$g_{\mu\nu}^{\pm}$ are continuous, which states that the
induced metrics from both sides $M_\pm,$ must be the same.
The condition originates from the fact that the derivative
of a step function is a delta function which generates a
singular term in the connection. The second junction
condition deals about the derivative of the metrics
$g_{\mu\nu}^{\pm}$. For a later convenience, we now
introduce such a useful notation that
\begin{equation}
[A] \equiv A^+|_{\Sigma_+} -A^-|_{\Sigma_-}
\end{equation}
where each $A^\pm$ is a tensorial quantities defined in
$M_\pm$. The second junction condition can be written as
\begin{equation}
[K_{ij}]=0
\end{equation}
where $K_{ij}$ is a extrinsic curvature which will be
defined below. In a coordinate system it is simply
proportional to the derivative of the metric. The second
junction condition plays a role of removing the
$\delta$-function terms from the Einstein equations and is
a sufficient condition for the regular behavior of the full
Riemann tensor at the hypersurface $\Sigma$. If this
condition is violated, then it means that the spacetime is
singular at $\Sigma$. We can cure this problem by the
following method. If the extrinsic curvature is not the
same on both sides of $\Sigma$, then a thin shell with the
$\delta$ function-like stress-energy tensor must be present
at $\Sigma$, which the junction equations tell us about.
This situation is similar to the eletrostaic problem
where a surface layer of nonzero electric charge density
exists between two different mediums. This sheet generates
a discontinuity in the electric field and a kink in the
electrostatic potential, which is completely analogous to
the gravitational case. These physical meanings can be
embodied more definitely in deriving the junction
equations, which starts from an action
\begin{equation}
S_{total} = S_{EH}+S_{GH}+S_{DW}
\end{equation}
where
\begin{eqnarray}
S_{EH} &=& \frac{1}{2\kappa^2}\int_M d^{D+1} x \sqrt{-g}~(R-2\Lambda),\\
S_{GH} &=& \frac{1}{\kappa^2}\int_{\Sigma_{\pm}} d^{D} x \sqrt{-h} K,\\
S_{DW} &=& \int_{\Sigma} d^{D} x \sqrt{-h} {\cal L}_{DW}.
\end{eqnarray}
We set $\kappa^2=8\pi G$ and $\Lambda$ cosmological
constants. $n^\rho$ is a unit normal vector and $h_{\mu\nu}
= g_{\mu\nu} - n_\mu n_\nu$ is the metric induced on
$\Sigma_{\pm}$. $K = h^{ij} K_{ij}$ is the trace of the
extrinsic curvature $K_{ij}= h^{\phantom{k} k}_i
h^{\phantom{k} l}_j \nabla_k n_l$ on $\Sigma_\pm$. $S_{EH}$
is the ordinary Einstein-Hilbert action. For a good
variational problem we must include the Gibbons-Hawking
action $S_{GH}$\cite{Gibbons:1976ue}. Varying $S_{EH}$ with
respect to each bulk metric generates boundary terms which
contain normal derivatives of the metric variations and
their presence will spoil the variational principle leading
to Einstein's equations. The terms from the metric
variations of $S_{GH}$ exactly cancel the boundary terms of
the normal derivatives of the metric in
$S_{EH}$\cite{Chamblin:1999ya}. Due to discontinuities of
$K_{\mu\nu}$ across $\Sigma$, the contributions from
$\Sigma_{\pm}$ need not cancel each other. Thus we need a
domain wall action. In general, when the gravitational
back-reaction of a domain wall is included, the global
causal structure of the resulting spacetime is usually
modified. For simplicity, we assume that no bulk fields do
not couple to the domain wall, whose world volume can be
described as a Nambu-Goto action
\begin{equation}
S_{DW}=-\sigma \int_{\Sigma}d^{D}x \sqrt{-h}
\end{equation}
where $\sigma$ is a parameter which corresponds to a
tension or energy density of the wall $\Sigma$. There can
be no transfer of energy between the bulk matter and the
wall because the energy density of the wall is fixed. In
this case, the stress-energy tensor of the domain wall is
proportional to the wall's induced metric tensor $h_{ij}$.
Now the metric variations of the total action $S_{total}$
give the Israel's junction equations
\begin{equation} \label{gj}
[K_{ij}] = \kappa^2\left(S_{ij}-\frac{1}{D-1}Sh_{ij}\right)
\end{equation}
where a stress tensor of the wall, tangential to the domain wall, is defined by
\begin{equation}
S_{ij} \equiv \frac{2}{\sqrt{-h}} \frac{\delta S_{DW}}{\delta h^{ij}}.
\end{equation}
With $S_{ij}=\sigma h_{ij}$, then the junction equations take a simple form
\begin{equation} \label{gj}
[K_{ij}] = -\chi_Dh_{ij}
\end{equation}
where
\begin{equation}
\chi_D=\frac{8\pi G_{D+1}\sigma}{D-1}.
\end{equation}
For applications of the junctions equations, we assume that
each bulk has a spherical symmetry such that the bulk
metric has the form
\begin{equation} \label{bulkmetric}
ds^2 = -H(R)dT^2+\frac{dR^2}{H(R)}+R^2d\Omega^2_{D-1}
\end{equation}
where
\begin{eqnarray}
H(R) &\equiv& 1-AR^2 - \frac{Bw_DG_{D+1}M}{R^{D-2}},\\
w_D &\equiv& \frac{16\pi}{(D-1)V_{D-1}},\\
V_{D-1}&\equiv&\frac{2\pi^{D/2}}{\Gamma(D/2)},
\end{eqnarray}
and $M$ is a black hole mass if any. The parameters
represent the followings : If a bulk resembles a black
hole-like one, $B=1$. Otherwise, $B$ is zero. The constant
$A$ is related with a cosmological constant, so that
$A=e/L^2=2e\Lambda/D(D-1)$ with $e=+1$ for dS space and
$e=-1$ for AdS space. We choose a positive direction so
that makes the radial coordinate with fixed angular
components increase. Due to the spherical symmetry, the
induced metric on the wall can be rewritten as
\begin{equation}
ds^2_{wall}=-d\tau^2+r(\tau)^2d\Omega^2_{D-1}
\end{equation}
where $\tau$ is the proper time of the world line of the
observer on the wall and satisfy the relation
\begin{equation}
d\tau^2=H(r)dt^2-\frac{1}{H(r)}dr^2
\end{equation}
with our convention that the bulk coordinates on the
position corresponding to the domain wall use lowercase
letters. We will interpret the radial coordinate of the
domain wall as a scale factor $r(\tau)$ from the viewpoint
of the wall observer's. At this point, we introduce the
Gauss normal coordinates near the wall $\Sigma$, which
require
\begin{equation}
g^{\eta \eta} =g_{\eta\eta}=1.
\end{equation}
In these coordinates, the extrinsic curvatures take very simple forms
\begin{eqnarray} \label{exc}
K^{\theta}_{\phantom{\mu}\theta}&=&K^{\phi}_{\phantom{\mu}\phi}=\cdots=\frac{\epsilon\sqrt{\dot{r}^2+H(r)}}{r}\\
K^{\tau}_{\phantom{\mu}\tau}&=&\frac{\epsilon}{\sqrt{\dot{r}^2+H(r)}}\left(\ddot{r}+\frac{1}{2}\frac{dH(r)}{dr}\right)
\end{eqnarray}
at the location of the wall $R=R(r(\tau))$. If the
spherical surface area increases with the radial
coordinate, $\epsilon=+1$ and $\epsilon=-1$ otherwise. The
dot represents a derivative with respect to the proper time
$\tau$ of an observer's worldline on the wall. Note that
the angular parts of the extrinsic curvature are
diagonalized and have the same value at each point of the
wall due to the spherical symmetry. Plugging (\ref{exc})
into (\ref{gj}) results in the equation
\begin{equation}
\epsilon_{out}\sqrt{\dot{r}^2+H_{out}(r)} - \epsilon_{in}\sqrt{\dot{r}^2+H_{in}(r)}= -\chi_D r,
\end{equation}
or
\begin{equation}
{\dot{r}}^2 - \frac{(H_{out}-H_{in})^2}{4\chi_D^2 r^2}
-\frac{1}{4}\chi_D^2 r^2 +\frac{1}{2}(H_{out} +H_{in}) = 0
\end{equation}
where the subscripts $out$(or $+$) and $in$(or $-$)
represent the outer side and the inner side of the wall
respectively. After a short algebra with the explicit forms
of $H_{\pm}$, the above equation can finally be rewritten
as
\begin{equation} \label{eff1}
\frac{1}{2}{\dot{r}}^2 + V(r) = -\frac{1}{2}
\end{equation}
where the effective potential is
\begin{eqnarray} \label{eff}
V(r)&=& -\frac{r^2}{8}\left[\frac{(A_{out}-A_{in})^2}{\chi_D^2}+
\chi_D^2 +2(A_{out}+A_{in})\right]\nonumber \\
&&-\frac{G_{D+1}w_{D}}{4r^{D-2}}\left[\frac{(A_{out}-A_{in})(B_{out}M_{out}-B_{in}M_{in})}{\chi_D^2}
+(B_{out}M_{out}+B_{in}M_{in})\right]\nonumber \\
&&  - \frac{G_{D+1}^2w^2_{D}(B_{out}M_{out}-B_{in}M_{in})^2}{8 \chi_D^2 r^{2D-2}}.
\end{eqnarray}
This equation of motion describes a fictitious particle
moving under the effective potential $V(r)$ with the energy
$-\frac{1}{2}$ analogous to a 1-dimensional classical
particle of energy E influenced by the potential $V$, which
we know how to deal very well. The remaining thing is
classifying the allowed solutions of the wall under the
given effective potential. The strategy is as follows. At
first, we start to find out the possible shapes of the
potential by collecting information about the extreme
values of the potential. The second, we simply compare the
extreme values of the potential with the energy $-1/2$. If
the maximum of the potential is always lower than $-1/2$,
then we call the motion of the wall as a `monotonic' motion
because it expands monotonically without bounds. If there
is a region in which the potential is higher than the
energy of the wall, the wall's motion will also be
restricted. In this case, if the wall has a
maximum(minimum) radius to which it expands, we call the
corresponding solution as a `bounded'(`bounce') solution
whose meaning is clear.

\section{Classifications of the wall motions}
Now, consider the junction equation describing the
connection between a maximally symmetric spacetime at
center and a black hole type spacetime in the outside
region. In general, the junction equation is given by
\begin{equation}   \label{const1}
\epsilon_{in} \sqrt{\dot{r}^2 + H_{in}} - \epsilon_{out}
\sqrt{\dot{r}^2 + H_{out}} = \chi_D r
\end{equation}
with
\begin{eqnarray}
H_{in} &=& - A_{in} r^2 +1 \nonumber \\
H_{out} &=& - A_{out} r^2 +1 - \frac{Bw_DG_{D+1}M}{r^{D-2}} .
\end{eqnarray}
Though the solution of the junction equation is obtained
from solving (\ref{eff}), we can find some constraints for
the region of $r$ from  (\ref{const1}). Since the tension
$\sigma$ is positive and $r$ runs form $0$ to $\infty$,
depending on the sign of $\epsilon$
the region of $r$ having a solution is restricted: \\
1) $\epsilon_{in} = \epsilon_{out} = 1$ \\
Having a solution, the position of the wall is located at the restricted region
$r^D < \frac{Bw_DG_{D+1}M}{A_{in} - A_{out} + \chi_D^2}$, where we set $A_{in} - A_{out}$
to a positive number. \\
2) $\epsilon_{in} = \epsilon_{out} = -1$\\
In this case, the region of the wall position is restricted as
$r^D > \frac{Bw_DG_{D+1}M}{A_{in} - A_{out} -\chi_D^2}$. \\
3) $\epsilon_{in} = - \epsilon_{out} = 1$\\
The wall is located at the region $\frac{Bw_DG_{D+1}M}{A_{in} - A_{out} + \chi_D^2} <r^D
< \frac{Bw_DG_{D+1}M}{A_{in} - A_{out} - \chi_D^2}$ in $r$-direction. \\
4) $- \epsilon_{in} = \epsilon_{out} = 1$\\
In this case, there is no region having a solution.

When solving (\ref{eff}), the above restricted region corresponding to each case
is also considered.

\subsection{(D+1)-dimensional M-SAdS Junction}

In this section we consider the M-SAdS type junction which is
described by the parameters
\begin{equation}
A_{out}=-\frac{2\Lambda_{out}}{D(D-1)}~,~B_{out}=1~,~M_{out}=M~,~
\textrm{and~ the~ others~ zero}.
\end{equation}
A positively-defined quantity $\chi_+$ is related to the vacuum
energy of the outer SAdS space. For convenience, we introduce the
ratios among the parameters by
\begin{equation}
u\equiv
\frac{1}{\chi_D}\sqrt{\frac{2\Lambda_{out}}{D(D-1)}}
~~\textrm{and}~~ v\equiv\frac{m}{(D-1)\chi_D}
\end{equation}
where
\begin{equation}
m=(D-1)w_DG_{D+1}M
\end{equation}
roughly corresponds to the mass density proportional to the black
hole mass divided by the volume of the unit sphere $S^{D-1}$. The
effective potential (\ref{eff}) becomes to
\begin{eqnarray} \label{MSAdS}
V(r)&=& -\frac{1}{8}(u^2-1)^2\chi_D^2r^2 +\frac{v(u^2-1)\chi_D}{4r^{D-2}}-\frac{v^2}{8r^{2D-2}}\nonumber\\
&=&-\frac{1}{8}\left((u^2-1)\chi_D r-\frac{v}{r^{D-1}}\right)^2
\end{eqnarray}
Therefore, the scheme of classifying the possible wall motion of
$M-SAdS$ junction can depend on the signs of $u$ and $v$.
\begin{figure} \label{fmsads}
\begin{center}
\includegraphics[height=.8\textheight, width=1\textwidth]{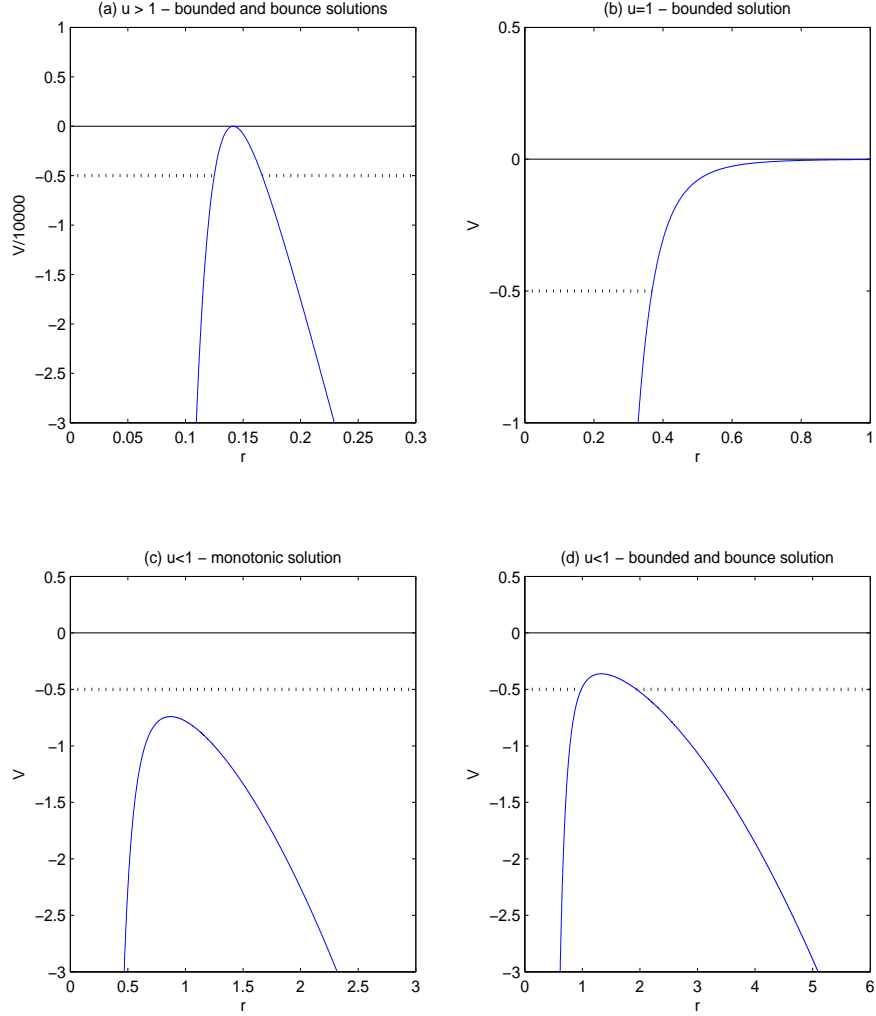}
\caption{Classification of the possible motions of the wall
in the 5 dimensional M-SAdS junction. We set $w_4G_5M=1$ in
all case for simplicity. (a) $u=50$ and $\chi_4=1$ (b)
$u=1$ and $\chi_4=10$ (c) $u=0.2$ and $\chi_4=1$ (d)
$u=0.4$ and $\chi_4=2.5$.}
\end{center}
\end{figure}

\subsubsection{$u>1$}
The value of the effective potential asymptotically diverges to
$-\infty$ as the radial coordinate goes to zero or infinity. This
means that there exists one critical point at which the potential
has a maximum value. The point is given by
\begin{equation}
r^D_c = \frac{v}{\chi_D (u^2-1)},
\end{equation}
at which the potential really has a maximum value equal to zero.
Therefore, the motions of the wall are limited to the bounded one or
the bounce one(Fig. 1(a)). In the bounce solution, the late time
behavior of the solution(large $r$) is
\begin{equation}
r \approx e^{\frac{\chi_D(u^2-1)}{2}\tau}
\end{equation}
independent of $m$.

\subsubsection{$u=1$}
\noindent The effective potential is
\begin{equation}
V(r) = - \frac{v^2}{8r^{2D-2}}.
\end{equation}
The only allowed solution of the junction equation is a bounded
type(Fig.1(b)), which can expand to a maximum radius
\begin{equation}
r_{max}=\left(\frac{v}{2}\right)^{\frac{1}{D-1}}
\end{equation}
and then shrink. In terms of $r_{max}$, the potential can be
rewritten by
\begin{equation}
V=-\frac{1}{2}\left(\frac{r_{max}}{r}\right)^{2D-2}
\end{equation}
and the trajectory $r=r(\tau)$ of the wall can be found by integrating
\begin{equation}
\pm \tau = \int_{r(0)}^{r(\tau)} \frac{dr
^{\prime}}{\sqrt{\left(\frac{r_{max}}{r ^{\prime}}
\right)^{2(D-1)}-1}}
\end{equation}
and getting the inverse of it. Around the origin the
behavior of the radius of the wall is
\begin{equation}
r \approx \left(Dr_{max}^{D-1}\cdot\tau\right)^{\frac{1}{D}}.
\end{equation}

\subsubsection{$u<1$}
The effective potential of this case resembles the shape of the
$u>1$ case. A extreme point of the potential occurs at
\begin{equation}
r^D_c = \frac{(D-1)v}{\chi_D(1-u^2)}.
\end{equation}
$V(r_c)$ corresponds to the maximum of the curve of the
effective potential and is always negative. Therefore, the
possible solutions can be classified by comparing $V(r_c)$
and the energy of the virtual one-particle. If
$V(r_c)<-1/2$, then a monotonic motion is possible(Fig.
1(c)). If $V(r_c)>-1/2$ , then a bounced or bounded motion
is allowed(Fig. 1(d)).

\subsubsection{Summary of the M-SAdS Junction}
We can classify the wall motions under the M-SAdS junctions
by $u$, $v$ and $\chi_D$. The extreme points $r_c$ are the
solutions of the ordinary second equation with respect to
the variable $r_c^D$. The classification corresponds to the
regions separated by surfaces in the 3-dimensional
parameter space $(\chi_D,u,v)$. For $u>1$, the allowed
solutions are bounded or bounce ones. For $u=1$, the
possible solution is limited to a bounded type. For $u<1$,
we need further information about $v$ and $\chi_D$ to
classify the wall motions completely in this parameters'
ranges. We found that all the 3-type motions are allowed in
that case.


\subsection{(D+1)-Dimensional AdS-SAdS Junctions}
We set
\begin{eqnarray}
H_{out} &=& 1+\frac{2\Lambda_{out}}{D(D-1)}R^2 -\frac{w_DG_{D+1}M}{R^{D-2}},\\
H_{in} &=& 1+\frac{2\Lambda_{in}}{D(D-1)}R^2.
\end{eqnarray}
For simplicity, we introduce another parameter ratio
\begin{equation}
w\equiv
\frac{1}{\chi_D}\sqrt{\frac{2\Lambda_{in}}{D(D-1)}}.
\end{equation}
According to (\ref{eff}), the effective potential is given by
\begin{equation}\label{padssads}
V(r)=-\frac{1}{8}\alpha\chi_D^2r^2 +\frac{\beta\chi_D
v}{4r^{D-2}} -\frac{v^2}{8r^{2D-2}}
\end{equation}
where
\begin{equation}
\alpha = \beta^2 - 4w^2 ~~\textrm{and}~~ \beta =u^2-w^2-1.
\end{equation}
Thus, the classification corresponds to divide the
4-dimensional parameter space $(\chi_D,u,v,w)$ into regions
by the judicious surfaces of the parameters. The regions
described by the curves $u>w$ in the $u-w$ plane represent
the junctions of the true-false vacuum type. Since we want
to concentrate on the false-true vacuum types, the regions
with positive $\beta$ are discarded. To see this, note that
$u=\pm w$ are asymptotes of the hyperbola $\beta=0$ in
$u-w$ plane. In this section, we have always the condition
$0<u<w$ in mind. We also set every parameters to be
positive. Of course, we could deal with the cases allowing
the negative ranges of the parameters but that simply
results in the mirror image with respect to the appropriate
axes or curves in the $u-w$ plane. In this paper, we ignore
the possibilities of those regions and the physical meanings of the negative range of the parameters.\\

\begin{figure} \label{adssads321}
\begin{center}
\includegraphics[height=.7\textheight, width=1\textwidth]{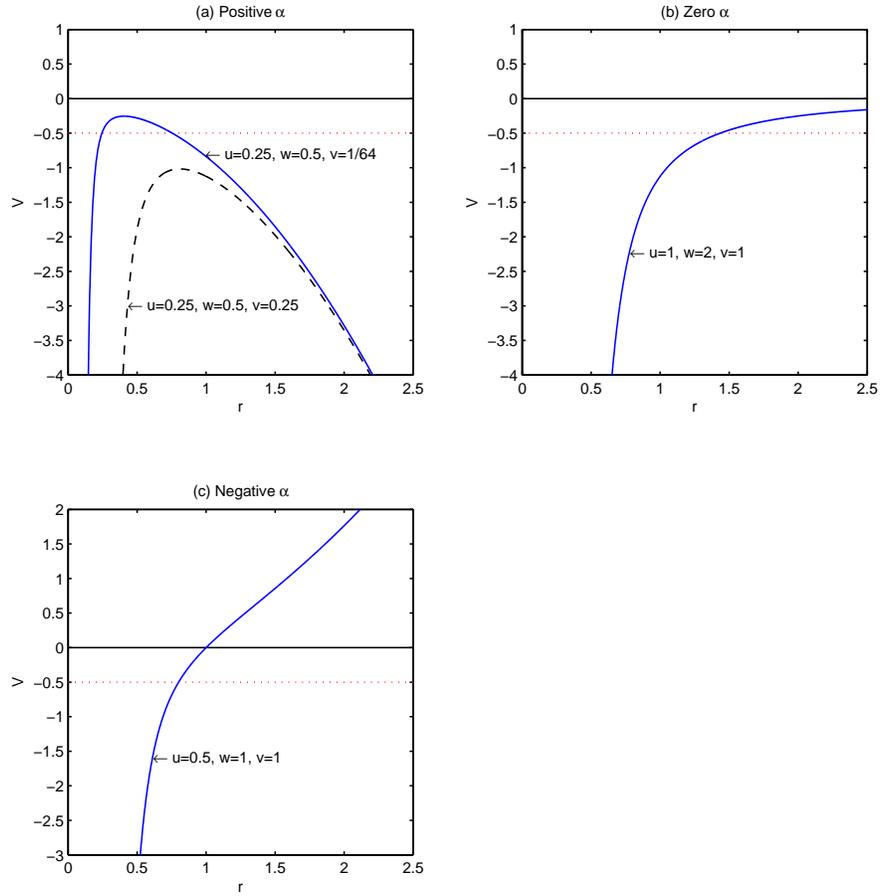}
\caption{Classifying the possible motions of the wall in the
AdS-SAdS junction by $\alpha$}
\end{center}
\end{figure}

\subsubsection{$\alpha > 0$}
The condition $\alpha>0$ is translated into $\beta<-2w$(Fig. 2(a)).
There are two regions in $u-w$ plane. The first region is
surrounded by
\begin{equation}
0<u<w ~~\textrm{and}~~ u+w-1<0
\end{equation}
and the second is enclosed by the curves
\begin{equation}
0<u<w ~~\textrm{and}~~ u-w+1<0.
\end{equation}
From the analysis of the asymptotic behaviors of the
effective potential, it is expected that the potential have
only one extreme point $r_c$. By an explicit calculation,
we find that it really does so that the extreme point is
\begin{equation}
r_c^D = \frac{v}{2\alpha\chi_D}\left[-(D-2)\beta
+\sqrt{(D-2)^2\beta^2+4(D-1)\alpha}\right],
\end{equation}
at which the potential has a maximum. The solutions in the
ranges of these parameters can be classified by comparing
the value of the maximum of the potential $V(r_c)$ and the
energy. If $V(r_c)$ is greater than the energy, the motion
of the wall can be either bounded or bounced type.
Otherwise, the possible motion of it should be a monotonic
one.

\subsubsection{$\alpha =0$}
The condition $\alpha=0$(Fig. 2(b)) corresponds to a straight line
\begin{equation}
u=\mid w-1\mid
\end{equation}
for $1/2<w$ except a point $(u,w)=(0,1)$. Therefore, two
different modes may exist. One corresponds to $w-u=1$ and
the other corresponds to $w+u=1$. From the analysis of the
potential, it is easy to see that there is only a bounded motion of the wall.\\

\subsubsection{$\alpha < 0$}
The negative $\alpha$ means that $\beta$ is in the range
\begin{equation} \label{ineq1}
-2w < \beta < 0,
\end{equation}
which corresponds to a strip in the $u-w$ plane surrounded by
\begin{equation}
0<u<w ~~\textrm{and}~~ u=\mid w-1\mid
\end{equation}
for $1/2<w$ except a point $(u,w)=(0,1)$(Fig. 2(c)).
The effective potential can be rewritten as
\begin{equation}
V(r)=\frac{1}{8}\mid\alpha\mid\chi_D^2r^2
+\frac{\beta\chi_D v}{4r^{D-2}} -\frac{v^2}{8r^{2D-2}},
\end{equation}
which is expected to have two positive real critical points
of the potential or no such ones to explain the asymptotic
behaviors of the potential naturally. Since the expressions
for two extremal points are given by
\begin{equation}
r_{c\pm}^D=\frac{v}{2\mid\alpha\mid\chi_D} \left[(D-2)\beta
\pm \sqrt{D^2\beta^2-16(D-1)w^2}\right],
\end{equation}
with a negative $\beta$, the potential can not have two
extreme points. We conclude that the motion of the wall is
always bounded in the range of negative $\alpha$.

\subsubsection{Summary of the AdS-SAdS Junctions}
The possible types of the motion of the wall is more
complicated than the ones of the previous section since we
turned on the negative cosmological constant inside the
wall and set the outer metric to be the form of the
Schwarzschild AdS type. We can classify the motions of the
wall simply as dividing the first quadrant in the $u-w$
plane into several regions according to the sign of
$\alpha$ and $\beta$. The monotonic type solution can only
exist in the case of $\alpha<0$.

\subsection{Classifying Scheme to the other (D+1)-dimensional Junctions}
In this section, we will apply our classifying scheme to
the other spherical false-true vacuum type of junctions.
The dS-S, dS-SdS, dS-AdS cases have been handled in many
other papers about the bubble dynamics and their possible
wall motions are a bounded or bounce or monotonic type
again. But their classifications dependent on the
parameters used are not given completely. As an exercise
for applying our scheme, we consider and try to classify
them again.

\subsubsection{(D+1)-dimensional dS-SdS Junctions}
To classify solutions of a junction, we must know what
effective potential it has. There are two ways that we can
do this. The one is that we simply assume the forms of the
bulk metrics as in the equation (\ref{bulkmetric}) and use
the formula (\ref{eff}) and the other is that we change the
signs of the combinatory parameters $u^2$ and $w^2$
appropriately. We follow the latter way in this
subsections. Given the effective potential, we can classify
the wall motions from a wall observer's view. The
coefficients of the bulk metrics in the dS-SdS junction are
\begin{equation}
A_{in}=w^2\chi_{D}^2 ~,~A_{out}=u^2\chi_{D}^2 ~,~M_{out}=M
~\textrm{and ~the~ others ~zero}.
\end{equation}
Because one of the differences between the dS space and the
AdS space is that they differ each other in the vacuum
energies, we flip the signs of $u^2$ and $w^2$ and substitute them in the AdS-SAdS case. Thus the
effective potential is given by
\begin{equation}\label{pdssds}
V(r)=-\frac{1}{8}\alpha\chi_D^2r^2 -\frac{\beta\chi_D
v}{4r^{D-2}} -\frac{v^2}{8r^{2D-2}}
\end{equation}
where
\begin{equation}
\alpha = \beta^2 + 4w^2 ~~\textrm{and}~~ \beta =u^2-w^2+1.
\end{equation}
Since the effective potential is given, we can classify the
solutions by the signs of $\alpha$ and $\beta$ and the expected forms of the potential. We omit the
detailed analysis of this junction.

\subsubsection{(D+1)-dimensional dS-AdS Junctions}
In this case, we can get the effective potential by
$u^2\rightarrow -u^2$ and $v=0$ in the potential of the
dS-SdS junction (\ref{pdssds}) or by $w^2\rightarrow -w^2$
and $v=0$ in the potential of the AdS-SAdS junction
(\ref{padssads}). The result is
\begin{equation}\label{pdsads}
V(r)=-\frac{1}{8}\alpha\chi_D^2r^2 -\frac{\beta\chi_D
v}{4r^{D-2}}
\end{equation}
where
\begin{equation}
\alpha = \beta^2 + 4w^2 ~~\textrm{and}~~ \beta =u^2+w^2+1.
\end{equation}

\subsubsection{(D+1)-dimensional dS-S Junctions}
This case simply corresponds to the $dS-SdS$ junctions with
replacing $u$ by zero, that is, we turn off
$\Lambda_{out}$. Thus the effective potential is
\begin{equation}\label{pdss}
V(r)=-\frac{1}{8}(1+w^2)^2\chi_D^2r^2 -\frac{\chi_D
v(1+w^2)}{4r^{D-2}} -\frac{v^2}{8r^{2D-2}}.
\end{equation}

\section{AdS - RN AdS  Junctions}
In this section we will deal with the junction equation which
connects the inner AdS space to the outer
Reissner-Nordstr$\ddot{o}$m AdS space. To describe the
Reissner-Nordstr$\ddot{o}$m AdS space, we consider the
charged particles spread uniformly on the domain wall. In
string theory set-up, the domain wall in the asymptotically
AdS space corresponds to the D3-brane and the charged
particles correspond not to the fluctuations of open string
on D3-brane but to D5-branes wrapped on $S^5$ in $AdS_5
\times S^5$. After some compatification mechanism, these
D5-branes look like charged particles in $AdS_5$ space and
become the sources of the bulk 1-form gauge field. So the
action describing this situation is
\begin{eqnarray}
S &=& \frac{1}{2\kappa^2}\int_M d^{D+1} x \sqrt{-g}\left(R- 2 \Lambda\right) -\frac{1}{4}\int_M d^{D+1}x\sqrt{-g}F^2 + \frac{1}{\kappa^2}\int_{\Sigma_{\pm}} d^{D} x \sqrt{-h} K \nonumber\\
&&  - \int_{\Sigma} d^{D} x \sqrt{-h} \sigma  + \int_{\Sigma} d^{D} x \sqrt{-h}  q C_0
\end{eqnarray}
with
\begin{equation}
F_{r0} = \partial_r C_0,
\end{equation}
where $q$ is a constant charge density on the domain wall
and the total charge $Q$ is given by $Q = \int_{\Omega_3}
d^3 x \sqrt{h'} q$ with the three sphere metric, $h'$. Note
that the tension $\sigma$ is the sum of that of D3-brane
and the energy density of charged particles. In the inner
region of the wall where there is no source, the solution
of the equation of motion for the gauge field is given by
$F=0$. Using this result, we can easily find that in the
inner region the pure AdS space becomes the solution of the
Einstein equation. In the outside region, the charged
particles gives the non-zero gauge field. In the asymptotic
region, due to the cosmological constant the metric is
given by another AdS-metric. If we choose the Gauss
surface, $\Gamma$  in this region and perform a volume
integration enclosed by Gauss surface, by the Stoke's
theorem we find a simple equation on this surface
\begin{equation}
\int_G d\Omega_3 R^3 F_{r0}= Q.
\end{equation}
Solving this equation, the gauge field $C_0$ is given by
$C_0 = - q/r^2$. When the observer in the outside region
feel a energy lump in center if $\Lambda_{in} > \Lambda_{out}$,
this implies that the metric of outside region is
Schwarzschild-type metric. Finally, the metric in outside
region is described by an AdS Schwarzschild-type metric
with a charge Q, which results in the AdS RN-type metric.
With all of these, we can set the metrics to
be
\begin{eqnarray}
H_{out} &=& 1+\frac{2\Lambda_{out}}{D(D-1)}R^2 -\frac{w_DG_{D+1}M}{R^{D-2}} +\frac{Q^2}{R^{2(D-2)}},\nonumber\\
H_{in} &=& 1+\frac{2\Lambda_{in}}{D(D-1)}R^2.
\end{eqnarray}
Under these, the junction equations are
\begin{equation}
\frac{1}{2}\dot{r}^2 + V = -\frac{1}{2}
\end{equation}
where the effective potential is
\begin{eqnarray}
V(r)&=&-\frac{1}{8}\alpha\chi^2_Dr^2 +\frac{\beta \chi_D v}{4r^{D-2}}
-\frac{\beta Q^2}{4r^{2(D-2)}}\nonumber\\
&&-\frac{v^2}{8r^{2(D-1)}} +\frac{vQ^2}{4\chi_Dr^{3D-4}}
-\frac{Q^4}{8\chi_D^2r^{2(2D-3)}}.
\end{eqnarray}
\begin{figure} \label{adsrnads4}
\begin{center}
\includegraphics[height=.3\textheight, width=1\textwidth]{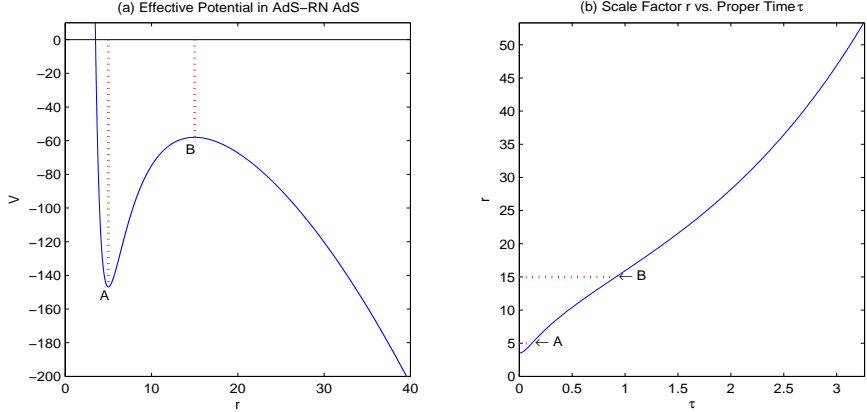}
\caption{An example of the junction between RN AdS and AdS
with multiple accelerations}
\end{center}
\end{figure}
Again, the possible motions of the wall can be classified
by the similar procedures as in the previous cases of Sec.
3. Here we only concentrate on a solution which can behave
like experiencing a multi-acceleration phase. From the form of
the induced metric on the wall, we may regard $r(\tau)$ as
a scale factor. Let the initial radius of the wall be
$r_0$. Refer to Fig. (3). When time goes on, the wall will expand. As the
radius of the wall approaches a point A at $r=5$, the scale factor behaves as $r\approx
\tau$ and the first acceleration, the so-called early inflation, ends. After that, the
expansion rate of the wall will be decelerated during some
time. After the wall goes over the point B at $r=15$, the
secondary acceleration will occur and the scale factor be
given by
\begin{equation}
r\simeq
\exp\left\{\frac{1}{2}\chi_D\sqrt{\mid\alpha\mid}\tau\right\}
\end{equation}
at $\tau=\infty$.

\section{Conclusions and Discussions}
In this paper, we classified all possible cosmological
behaviors from the viewpoint of an observer on the
domain wall moving in 5-dimension. The wall plays a
role of a partition that separates the whole manifold into
two bulks $M_\pm$ and acts on each bulk as a boundary.
The junction equation describes the motion of the wall in the radial
coordinate and can be interpreted as an equation for the
energy conservation of a classical particle under the
special potential. Physically, this equation is related
with the cosmology of the domain wall because the coordinate $r$ is considered as a scale factor of the metrics of the wall.

It is not clear whether the outside bulk metric can take the form of SAdS in true-false vacuum junction. Hence, we considered only false-true vacuum junctions. The classifications were
possible from the shapes of the effective
potential depending on the parameter combinations $\alpha$ and
$\beta$. From the equation $dV/dr=0$, the extremal
points were given by the solution of the quadratic
equations of $r_c^D$. As an example of these, we considered
M-SAdS and AdS-SAdS types in detail and their
classifications were given by dividing the involved
parameter spaces into the corresponding regions. For the dS-S, dS-SdS and dS-AdS junctions, we can get the junction equations by a judicious choice of the parameters.

Finally, we considered a possibility whether solutions with
two or more accelerations could exist and found out a special
example in the AdS-RN AdS junctions numerically. Without the full classification, we obtained the solution
with  the late time acceleration as well as the early time inflation. The reason why this was possible was
roughly due to the existence of the several extremal
points. Since many parameters are used, the positions and the times of the
inflations may be adjusted safely. Given a junction, if the effective potential of the problem has many extremal points, it will be possible for a model in the junction to give rise to multiple accelerations.

\vspace{1cm}

{\bf Acknowledgement} We thank E. J. Weinberg for his valuable discussion. This work was supported by the
Science Research Center Program of the Korea Science and
Engineering Foundation through the Center for Quantum
Spacetime(CQUeST) of Sogang University with grant number
R11 - 2005 - 021.

\vspace{1cm}


\end{document}